\documentclass[
twocolumn,showpacs,aps,prd]{revtex4}
\usepackage{dcolumn}
\usepackage{amsthm,amsmath,amssymb}
\usepackage{amsfonts,bm,latexsym}

\def\be{\begin{equation}}
\def\ee{\end{equation}}
\def\bea{\begin{eqnarray}}
\def\eea{\end{eqnarray}}

\def\nn{\nonumber}
\def\p{\partial}

\begin{document}


\title{Approach of background metric expansion to a new metric ansatz
for gauged and ungauged Kaluza-Klein supergravity black holes}

\author{Shuang-Qing Wu\email{sqwu@phy.ccnu.edu.cn, Corresponding author}
and He Wang\email{herbwang1989@gmail.com}}
\affiliation{Institute of Theoretical Physics, China West Normal University, Nanchong,
Sichuan 637002, People's Republic of China}


\begin{abstract}
In a previous paper [S.Q. Wu, Phys. Rev. D \textbf{83}, 121502(R) (2011)], a new kind of metric
ansatz was found to fairly describe all already-known black hole solutions in the ungauged
Kaluza-Klein (KK) supergravity theories. That metric ansatz somewhat resembles to the famous
Kerr-Schild (KS) form, but it is different from the KS one in two distinct aspects.
That is, apart from a global conformal factor, the metric ansatz can be written as a vacuum
background spacetime plus a ``perturbation" modification term, the latter of which is associated
with a timelike geodesic vector field rather than a null geodesic congruence in the usual KS
ansatz. Replacing the flat vacuum background metric by the (anti)-de Sitter [(A)dS] spacetime,
the general rotating charged KK-(A)dS black hole solutions in all higher dimensions have been
successfully constructed and put into a unified form. In this paper, we shall study this novel
metric ansatz in detail, aiming at achieving some inspiration as to the construction of
rotating charged AdS black holes with multiple charges in other gauged supergravity theories.
We find that the traditional perturbation expansion method often successfully used in the KS
form is no longer useful in our new ansatz, since here no good parameter can be chosen as a
suitable perturbation indicator. In order to investigate the metric properties of the general
KK-AdS solutions, in this paper we devise a new effective method, dubbed the background metric
expansion method, which can be thought of as a generalization of the perturbation expansion method,
to deal with the Lagrangian and all equations of motion. In addition to two previously known
conditions, namely the timelike and geodesic properties of the vector, we get three additional
constraints via contracting the Maxwell and Einstein equations once or twice with this timelike
geodesic vector. In particular, we find that these are a simpler set of sufficient conditions
to determine the vector and the dilaton scalar around the background metric, which is helpful
in obtaining new exact solutions. With these five simpler equations in hand, we rederive
the general rotating charged KK-(A)dS black hole solutions with spherical horizon topology
and obtain new solutions with planar topology in all dimensions. It turns out that the overall
calculations in finding the solution to the KK gauged supergravity can be reduced considerably,
compared to the previous process, by directly solving all the field equations. It is then shown
that the rotating charged KK-AdS black hole solutions can be further generalized by introducing
one or two arbitrary constants, while the black hole solutions with the planar AdS background
metric in all higher dimensions are newly obtained.
\end{abstract}


\maketitle

\section{Introduction}

It has been known for a long time that Einstein's gravitational field equations are such a very
complicated system of nonlinearly coupled partial differential equations that finding a rotating
exact solution to them is rather difficult. One of the frequently used approaches to this problem
is to assume an appropriate metric form for the unknown line element which is inspired from a
known solution in order to simplify the subsequent calculations. A well-known example of this
is provided by the Kerr solution \cite{Kerr}, which can be cast into the famous Kerr-Schild (KS)
form \cite{KS65,KS09,KS10}
\be
g_{ab} = \eta_{ab} +2H\, k_ak_b \, , \label{KSg}
\ee
where the vector $k^a$ is null and geodesic congruence with respect to the flat background metric
$\eta_{ab}$, and $H$ is a scalar function. Due to the fact that many interesting properties
are shared by this family of the metric ansatz \cite{ExactSols} and their applications result
in a substantial simplification of the field equations, a variety of generalizations of the KS
metric form have been accomplished during the past decades. Below we present a brief outline
of the main developments that have been achieved on the generalizations of the Kerr-Schild metrical
structure.

(a) The generalized Kerr-Schild ansatz, in which the background metric is replaced by arbitrary
spacetimes ($\eta\rightarrow \bar{g}$), was proposed and analyzed in Refs. \cite{Taub,MaSe}.
Many previous studies have shown that the Einstein field equations become linear within
the generalized Kerr-Schild ansatz for vacuum \cite{Xanthopoulos1} and nonvacuum
\cite{Xanthopoulos2,Gergely} spacetimes.

(b) The double Kerr-Schild metric, namely
\be
g_{ab} = \bar{g}_{ab} +2P\, k_ak_b +2Q\, l_al_b +4R\, k_{(a}l_{b)}
\ee
with $\bar{g}_{ab}k^ak^b = \bar{g}_{ab}l^al^b = \bar{g}_{ab}k^al^b = 0$, where $P, Q$, and $R$ are
three scalar functions, was introduced \cite{Plebanski1,Plebanski2,Plebanski3,Hall,McIntosh}
in the context of complex relativity in dimension $D = 4$. It is often misunderstood that in
Lorentzian signature, two such null vectors must be proportional. However, this is not always
the case if one considers the \emph{complex} Riemannian space with Lorentzian signature. To
explain this point, an explicit example for the four-dimensional Kerr-Newman-Unti-Tamburino-AdS
(Kerr-NUT-AdS) metric is provided in the Appendix. Further observations indicate that such a
general class of metric ansatz could be further generalized to the double Kerr-Schild metric
\cite{doubleKS} in $D > 4$ and the multi-Kerr-Schild metric \cite{mulKS} in $D\geqslant4$, where
the orthogonality properties still hold.

(c) An earlier generalization of the Kerr-Schild ansatz by introducing a nonzero cosmological
constant in four-dimensional spacetimes was considered in \cite{Kowalczy}. Further studies
\cite{ConformalKS1} showed that all static spherically symmetric vacuum spacetimes with or
without the cosmological constant can be described by conformal Kerr-Schild metrics
\cite{ConformalKS2}.

(d) Higher-dimensional generalization of the Kerr-Schild metric was firstly utilized by Myers
and Perry \cite{MP86} to construct exact, asymptotically flat vacuum solutions of rotating
black holes in all higher dimensions $D > 4$. Recently, rotating vacuum black holes with
a nonzero cosmological constant in higher dimensions were successfully constructed in
\cite{GLPP04,GLPP05} by simply replacing the flat background metric of the higher-dimensional
KS form by the pure (A)dS spacetime. Moreover, further investigations have demonstrated that
the background metric can be replaced by other asymptotically, locally flat spacetimes such
as those with the NUT charges; an important example of this includes the NUT extension
\cite{CLP06b} of rotating black holes in (A)dS spacetimes. General properties of higher-dimensional
Ricci-flat and (A)dS Kerr-Schild metric mentioned above were studied recently in
\cite{Ortaggio1,Ortaggio2,Malek1,Malek2}.

(e) One recent extension of the original KS form is named the extended Kerr-Schild ansatz
in \cite{CCLPxks,xKS}, where the metric of rotating charged black hole, namely the CCLP
\cite{CCLP} spacetime found in $D = 5$ minimal gauged supergravity, can be redescribed
in the framework of Kerr-Schild formalism as
\be
g_{ab} = \bar{g}_{ab} +H\, k_a k_b +V\, (k_al_b +l_ak_b)\, ,
\ee
in which H and V are two scalar functions to be specified; $k^a$ is again a null vector, and
the vector $l^a$ is spacelike and orthogonal to $k^a$ with respect to $\bar{g}_{ab}$, a flat or
(A)dS background metric.

The general properties such as geodesic and optical properties of the null congruence and Weyl
types of this kind of metric have been investigated in \cite{Malek2014,MalekPhD}. In addition,
it is worth pointing out that a limited case for the related extension of the Kerr-Schild
ansatz was studied in \cite{Bonanos}, where the background is Ricci flat in $D = 4$.

\medskip
In our previous paper \cite{SQWu}, a new kind of metric ansatz was found to satisfactorily
describe all already-known black hole solutions in the ungauged Kaluza-Klein (KK) supergravity
theories, which can be written in a new unified form in all higher $D > 4$ dimensions by
\be\begin{split}
g_{ab} & = H^{\frac{1}{D-2}}\Big(\bar{g}_{ab} +\frac{2m}{UH}k_{a}k_{b}\Big) \, , \\
g^{ab} & = H^{\frac{-1}{D-2}}\Big(\bar{g}^{ab} -\frac{2m}{U}k^{a}k^{b}\Big) \, ,
\end{split}\ee
where the background metric is a flat one. This metric ansatz is different from the entire
above-mentioned generalization of the original KS ansatz. It somewhat resembles
to the famous Kerr-Schild form, but there are significant differences from the KS one in
two distinct aspects, that is, apart from a common conformal factor, the vector $k^a$ is no
longer null but now it is timelike with respect to the background metric. The timelike vector
field $k^a$ is geodesic and its norm with respect to the background metric depends on the charge
parameter: $\bar{g}_{ab}k^ak^b = -s^2$. It should be noticed that in the uncharged case, the
conformal factor becomes unity and the vector $k^a$ becomes null; then our new metric ansatz
exactly reduces to the original KS metric (\ref{KSg}).

It has further been observed in Ref. \cite{SQWu} that one can adopt the pure (A)dS spacetimes
as the background metric and find the general rotating charged Kaluza-Klein (A)dS black hole
solutions with a single electric charge and arbitrary angular momenta as the exact solutions
to the Einstein-Maxwell-dilaton theory described by the following Lagrangian ($\mathcal{F} = dA$)
\bea
&&\hspace{-8pt} \mathcal{L} = \sqrt{-g}\Big\{R -\frac{1}{4}(D-1)(D-2)(\p\Phi)^2
 -\frac{1}{4}e^{-(D-1)\Phi}\mathcal{F}^2 \nn \\
&&\quad +g^2(D-1)\big[(D-3)e^{\Phi} +e^{-(D-3)\Phi}\big]\Big\} \, . \label{sKSl}
\eea
For the sake of later simplicity, the spacetime of this general form shall be briefly called
the stringy Kerr-Schild or sKS metric, since its metric structure has some relation to the
well-known Kerr-Schild form, and it is universal for almost all of the charged black hole solutions
already known in gauged supergravity theories. As such, the underlying metric structure of
our sKS form can be thought of as the most meaningful generalization of the Kerr-Schild ansatz
until now. In addition, it should also be mentioned that the $D = 4$ sKS metric can be expressed
as a form similar to those proposed by Yilmaz \cite{WCdS11} and later by Bekenstein \cite{JDB04}.
In addition, many studies have brought out that the Gordon metric can be further applied to
massive and bimetric theory (see \cite{Gordon1,Gordon2,Gordon3}), the metric structure of
which also resembles the generalized sKS metric form specifically in $D = 4$.

The main subject of this paper is to investigate the general properties of the field equations
for the sKS metric, since such an important theoretical analysis had not been delivered before
in any previous work. Moreover, our motivation of this study is to see whether the results of
such a theoretical analysis could be helpful in obtaining new exact solutions with the help of
the sKS form and in achieving some insights on constructing exact solutions to other gauged
supergravity theories.

The organization of this paper is outlined as follows. To begin with, in Sec. II we will show
that it is infeasible to analyze the sKS ansatz by the usual method of perturbation expansion.
Although this method proved to be inappropriate for our aim, one can still get a little
inspiration from it. As a replacement, we therefore put forward a new method, named the
background metric expansion, which can be viewed as the generalization of the previous one.
In the actual derivation, it is much more cumbersome for the sKS metric to express the Ricci
tensor and field equations, so it is convenient and in fact necessary to use the computer algebra
system Cadabra \cite{Cadabra1,Cadabra2}, which allows us to perform analytically symbolic
calculations in arbitrary unspecified dimensions. In Sec. III, our new method is used to obtain
the geodesic property of the vector $k^a$ with respect to the background metric, and a simpler
set of sufficient conditions of the field equations around the background metric are deduced.
With these results in hand, in Sec. IV we assume a vector $\tilde{K}$ to be timelike and geodesic
and then solve the simpler set of field equations around the pure AdS background spacetime.
Consequently, we obtain an extended version of rotating charged KK-AdS solutions with one or
two arbitrary constants (one for $\epsilon = 1$, two for $\epsilon = 0$). As another verification
of the effect of our method, we further obtain new exact solutions with planar topology by replacing
the background metric as a planar AdS metric \cite{Awad}. In Sec. V, we summarize our results and
discuss the prospect of the analysis, for which it would be very helpful for finding new exact
black hole solutions.

\section{Infeasibility of traditional
perturbation method for the sKS ansatz}

In this section, we present the usual formalism of perturbation expansion that goes into these
calculations for analogy and explain why the ordinary procedure is limited and inappropriate
for our new sKS form.

For simplicity, we use the dilaton scalar $\Phi$ to reexpress the metric tensors and the gauge
potential as \cite{SQWu}
\be\begin{split}
g_{ab} &= e^{-\Phi}\bar{g}_{ab} +\lambda\big[e^{-\Phi} -e^{(D-3)\Phi}\big]k_{a}k_{b} \, , \\
g^{ab} &= e^{\Phi}\bar{g}^{ab} +\lambda\big[e^{\Phi} -e^{-(D-3)\Phi}\big]k^{a}k^{b} \, , \\
A_{a} & = \sqrt{\lambda}\big[1 -e^{(D-2)\Phi}\big] k_{a} \, , \quad
 \Phi = \frac{-1}{D-2}\ln(H) \, ,
\end{split}\label{sKSlambda}
\ee
where the vector $k^a$ is a timelike geodesic congruence with respect to the AdS background
metric $\bar{g}_{ab}$ and satisfies $k_a = \bar{g}_{ab}k^b$, $k_ak^a = \bar{g}_{ab}k^ak^b = -1$.
The $\lambda$ is inserted here as a dimensionless parameter that would take the value $\lambda = 1$
finally. When $\lambda = 1$, we have for the full metric tensor the following properties:
\be\begin{split}
& g_{ab}k^b = e^{(D-3)\Phi}k_{a} \, , \quad g_{ab}k^ak^b = -e^{(D-3)\Phi} \, , \\
& g^{ab}k_b = e^{(3-D)\Phi}k^{a} \, , \quad g^{ab}k_ak_b = -e^{(3-D)\Phi} \, .
\end{split}\ee

We note that the timelike vector $k^a$ also satisfies the geodesic property: $k^a\bar{\nabla}_a
k^b = 0$ with the specific spacetime, where $\bar{\nabla}_a$ denotes the covariant derivative
operator compatible with the background metric $\bar{g}_{ab}$. Although this geodesic property
is very helpful to simplify our computations during the subsequent perturbation process, we
will only apply the essential relations (\ref{sKSlambda}) and the timelike condition to do
perturbational analysis in an attempt to figure out the most universal properties of the sKS
ansatz.

The curvature of the sKS metric as well as other useful quantities can be computed in terms of
the curvature of the background metric $\bar{g}_{ab}$ and the background covariant derivative
of the vectors $k^a$. The action of the full covariant derivative on a vector can be written
as $\nabla_{a}v^b = \bar{\nabla}_a v^b +C^b{}_{ac}v^c$, in which the connection $C^{c}_{~ab}$
is given by
\be
C^{c}_{~ab} = \frac{1}{2}g^{cd}\big(\bar{\nabla}_a g_{bd} +\bar{\nabla}_b g_{ad}
 -\bar{\nabla}_c g_{ab} \big) \, ;
\ee
then the Ricci tensor of $g_{ab}$ is related to that of $\bar{g}_{ab}$ by
\be
R_{ab} = \bar{R}_{ab} -2\bar{\nabla}_{[a}C^c_{~c]b} +2C^e_{~a[b}C^c_{~c]e} \, .
\ee
(See also \cite{ExactSols}.) The determinant of the full metric of sKS form is related to the
background one by $\sqrt{-g} = e^{\Phi}\sqrt{-\bar{g}}$; hence we have an identity $C^b{}_{ab}
= -\bar{\nabla}_a\Phi$.

After using Cadabra \cite{Cadabra1,Cadabra2} software to undertake the tedious calculations,
we can write the connection coefficients and the Ricci tensor containing terms quadratic in
connection coefficients as a sum over contributions at different powers in $\lambda$ as follows:
\be\begin{aligned}
C^{c}_{~ab} &= \sum_{k=0}^{2}\lambda^k C^{c(k)}_{~ab} \\
 &= \frac{1}{2}\bar{g}_{ab}\bar{\nabla}^c\Phi -\bar{g}^c_{~a}\bar{\nabla}_b \Phi
  +\sum_{k=1}^{2}\lambda^k C^{c(k)}_{~ab} \, , \\
R_{ab} &= \sum_{l=0}^{4}\lambda^l R^{(l)}_{ab} \, . \label{Rab}
\end{aligned}\ee
Based upon these expressions, the Einstein equation $E_{ab} = 0$ for the KK-AdS spacetime
(\ref{sKSl}) in terms of the background metric $\bar{g}_{ab}$ can be represented as $E_{ab}
= \sum_{n=0}^{4}\lambda^n E^{(n)}_{ab}$. Note that, in the uncharged case ($\Phi = 0 = A_a$),
the metric structure (\ref{sKSlambda}) and all the above equations reduce to the original
Kerr-Schild form.

In Eq. (\ref{Rab}), we have only considered the expansion of the Ricci tensor $R_{ab}$ in terms
of the parameter $\lambda$. One can also work with the mixed tensor $R^a_{~b}$. Unlike the case of
the standard Kerr-Schild form, the tensor $R^a_{~b}$ now still contains nonlinear terms in $\lambda$,
just like $R_{ab}$. This is because the vector $k^a$ is a timelike, rather than a null vector. Since
both $R_{ab}$ and $R^a_{~b}$ contain nonlinear terms in $\lambda$, it is of no priority to consider
the component with the mixed indices. For this reason, we prefer considering the full covariant
component in this paper. If one would like to work with the mixed components, then it is easy to
find that they are just a recombination of the covariant components.

To proceed further, it is facilitated by directly considering the contracted equation $E_{ab}
k^ak^b = 0$. Now one would naively expect the corresponding expressions for $E^{(n)}_{ab}k^ak^b$
in front of $\lambda^n$ at each order to vanish just as in the previous case considered in
\cite{xKS}. Our computation shows that the contribution from the fourth order $E^{(4)}_{ab}
k^ak^b$ vanishes identically, while at order $\lambda^3$ it reads
\be \begin{aligned}
E^{(3)}_{ab}k^ak^b=& -\frac{(1-\gamma)^4}{2(D-2)\gamma^2}(\bar{D}k_a)\bar{D}k^a \\
& +\frac{3\gamma -5 +D}{2}\big(1 -\gamma^{-1}\big)^2(\bar{\nabla}^a\phi)\bar{D}k_a \\
& +\frac{\alpha}{4\gamma^2}(1 -\gamma)\big[\bar{D}\Phi+(\bar{\nabla}\Phi)^2\big] \, ,
\label{E3}
\end{aligned}
\ee
where we denote $\gamma = e^{-(D-2)\Phi}$ and $\alpha = \gamma^2 +2(D-5)\gamma +D^2 -8D +13$,
while $\bar{D} = k^a\bar{\nabla}_a$ is the background covariant derivative taken along the null
vector $k^a$. Obviously, once we consider the geodesic property $\bar{D}k_a = 0$, the expression
(\ref{E3}) vanishes identically if and only if the condition $\bar{D}\Phi +(\bar{\nabla}\Phi)^2 = 0$
is satisfied, where we have taken a nonzero scalar function $\Phi$ into account (and thus $\gamma\neq
1$, $\alpha\neq 0$, and $3\gamma \neq D -5$). However, it is clear to check that this condition is
inconsistent with the explicit KK-AdS solutions in \cite{SQWu}, indicating that this is a meaningless
condition.

We have also attempted to place the perturbation factor $\lambda$ in different positions of the
metric ansatz, but still failed to move forward. Actually, it not only shows that the insertion
of the perturbation factor $\lambda$ is not suitable, but also reveals that this insertion is
irrational and unreasonable for the sKS ansatz. In fact, the underlying reason is that there
exists no perturbation parameter (neither the mass nor the charge) as an appropriable indicator
to do the corresponding perturbation analysis here, unlike the cases that are successful for
the original KS form and the extended Ks 
ansatz, where the mass parameter can be treated as a perturbation parameter. Due to the feature
of the conformal factor structure of the sKS metric ansatz, obviously direct application of the
ordinary perturbation expansion fails for the present situation; therefore, as an alternative
program, a new analysis method is needed.

\section{New method of background metric
expansion for the sKS metric ansatz}

In this section, we will now propose a new background metric expansion method towards analyzing
the sKS ansatz for the Einstein-Maxwell-dilaton system, which can be seen as a generalization
of the ordinary perturbation expansion method. In this new method, we will synthetically consider
the entire expansions of the Lagrangian and all the field equations around the background spacetimes,
in terms of the background metric and the background covariant derivatives, not just from the
viewpoint of perturbation expansion by which each term can be sorted in terms of the different
orders of $\lambda$. We then contract the Maxwell and Einstein equations once or twice with the
timelike vector $k^a$ to extract more useful information. Here, we are interested in seeing what
simplifications will occur and, in particular, what the implications of the resulting Lagrangian
and field equations will make. From the alternative perspective, we would like to see whether the
vector $k^a$ and the scalar $\Phi$ satisfy some conditions that could be helpful for us to obtain
new exact solutions. In doing so, we find that in addition to two previously known timelike and
geodesic properties obeyed by the vector, one can get three additional constraint equations.

\subsection{The Lagrangian expanded
around the background metric}

As shown in the previous section, there is no use in treating the $\lambda$ as a perturbation
parameter in the sKS metric ansatz; therefore, in the following we shall take $\lambda = 1$. Fortunately,
one can still expand all expected quantities in terms of those of the background metric. Our staring
point is the sKs ansatz (\ref{sKSlambda}) with $\lambda = 1$. With the help of Cadabra, the explicit
expression of the Ricci scalar is given in terms of the background metric and the background covariant
derivatives as
\bea\begin{aligned}
R = e^{\Phi} & \Big\{\bar{R} +\bar{\square}\Phi
 +(1-\gamma)\big[\bar{R}_{ab}k^ak^b -\bar{\nabla}^a\bar{\nabla}^b(k_ak_b)\big] \\
& -\frac{1}{4}(D-2)(D-3) (\bar{\nabla}\Phi)^2 \\
& +[1 +(1-D)\gamma] k_ak_b \bar{\nabla}^a\bar{\nabla}^b\Phi \nn \\
& +\frac{1}{2\gamma}(1-\gamma)^2\big[2(\bar{\nabla}^{a}k^{b})\bar{\nabla}_{[a}k_{b]}
 +(\bar{D}k_a)\bar{D}k_b \big] \\
& +[3-D +(1-D)\gamma](\bar{D}k_a)\bar{\nabla}^a\Phi \\
& +[1 +(3-2D)\gamma](\bar{D}\Phi)\bar{\nabla}_ak^a \\
& +\frac{1}{4}(D-2)[3-D +3(D-1)\gamma](\bar{D}\Phi)^2 \Big\} \, .
\end{aligned} \eea
Continuing to calculate the Lagrangian expanded around the background metric, we could obtain the
Lagrangian of the Einstein-Maxwell-dilaton system with respect to the background metric
\be\begin{aligned}
\mathcal{L} = \sqrt{-\bar{g}}
&\Bigm\{ \bar{R} +(1-\gamma)\big[\bar{R}_{ab} k^a k^b -(D-1)g^2\big] \\
& +g^2(D-1)(D-2) +\frac{1}{2}(1-\gamma)^2(\bar{D}k_a)\bar{D}k^a \\
& +\bar{\nabla}^a \big([1 +(1-D)\gamma]k_a\bar{D}\Phi \big) \\
& -\bar{\nabla}^a \big[(1-\gamma)\bar{\nabla}^b(k_ak_b)\big] \Bigm \} \, ,
\end{aligned} \label{LLL}
\ee
in which the relation $\sqrt{-g} = e^{\Phi}\sqrt{-\bar{g}}$ has been used.

Now we assume the background metric $\bar{g}_{ab}$ is the pure AdS metric and substitute $\bar{R}_{ab}
= -g^2(D-1) \bar{g}_{ab}$ into the above expression; then we get
\be\begin{aligned}
\mathcal{L} = \sqrt{-\bar{g}}
& \Bigm\{ \bar{R} +g^2(D-1)(D-2) \\
& +\frac{1}{2}(1-\gamma)^2(\bar{D}k_a)\bar{D}k^a \\
& +\bar{\nabla}^a \big([1 +(1-D)\gamma]k_a\bar{D}\Phi \big) \\
& -\bar{\nabla}^a \big[(1-\gamma)\bar{\nabla}^b(k_ak_b) \big] \Bigm \} \, .
\end{aligned} \label{PertL}
\ee
This expression establishes the association between the Lagrangian for the full metric and that of the
background metric. The variation of this Lagrangian with respect to the timelike vector $k_a$ implies
that $\bar{D}k^a$ must be a null vector: $(\bar{D}k_a)\bar{D}k^a = 0$. Since $\bar{D}k^a$ is also
orthogonal to the timelike vector $k^a$, without loss of generality we can take
\be
\bar{D}k^a = 0 \, . \label{geoc}
\ee
This is equivalent to the statement that $k^a$ is tangent to an affinely parameterized 
timelike geodesic congruence of the background metric. Besides, one can also arrive at the same conclusion
by solving vector $k^a$ within the field equations derived from the Lagrangian (\ref{PertL}) with respect
to $\bar{g}^{ab}$ and $\Phi$. In the following, we will proceed to consider the expansions of all the
equations of motion derived directly from the Lagrangian (\ref{sKSl}) by assuming that the background
spacetime is the pure AdS metric and vector $k^a$ is tangent to a congruence of affinely parameterized
timelike geodesics with respect to the background metric to find out more properties or relations for
further research.

\subsection{Field equations expanded
around the background metric}

We now turn to consider all the field equations deduced directly from the Lagrangian (\ref{sKSl}) and expand
them around the background metric. Calculating the variational derivatives of the Lagrangian (\ref{sKSl})
with respect to ($g^{ab}, A_a, \Phi$), one can obtain the contracted Einstein equation
\bea\begin{aligned}
& R_{ab} -\frac{1}{4}(D-1)(D-2)(\nabla_{a}\Phi)\nabla_{b}\Phi \\
&\qquad -\frac{\gamma}{2}e^{-\Phi}\Big[F_{ac}F_{bd}g^{cd} -\frac{g_{ab}}{2(D-2)}F^2\Big] \\
&\qquad +g^2\frac{D-1}{D-2}(D-3 +\gamma)e^\Phi g_{ab} = 0 \, ,
\end{aligned} \label{ree}
\eea
while the dilaton and gauge field equations are
\bea\begin{aligned}
\square\Phi +\frac{\gamma e^{-\Phi}}{2(D-2)}F^2 & +2g^2\frac{D-3}{D-2}(1-\gamma)e^{\Phi} = 0 \, , \\
& \nabla^a\big[\gamma e^{-\Phi} F_{ab}\big] = 0 \, .
\end{aligned} \label{rfe}
\eea

To avoid the clutter 
expressions, we first introduce the following two notations:
\bea
\bar{V}_a &\equiv& -2\bar{\nabla}^a\big\{(1-\gamma)\bar{\nabla}_{[a} k_{b]}
 +(D-2)\gamma(\bar{\nabla}_{[a}\Phi) k_{b]} \nn \\
&& +(1-\gamma)^2k_{[a}\bar{D}k_{b]}\big\} \, , \\
\bar{S} &\equiv& e^{\Phi} \Big\{\bar{\square}\Phi
 +\frac{2\gamma(1-\gamma)^2}{D-2}(\bar{\nabla}^ak^b)\bar{\nabla}_{[a}k_{b]} \nn \\
&& -(D-2)(\bar{\nabla}\Phi)^2 +\frac{\gamma^{-1}}{D-2}(1-\gamma)^3(\bar{D}k^a)\bar{D}k_a \nn \\
&& +(1-\gamma)\big[\bar{\nabla}^a(k_a\bar{D}\Phi) -(D-2)(\bar{D}\Phi)^2 \nn \\
&& -(\bar{D}k_a)\bar{\nabla}^a\Phi\big] +2g^2\frac{D-3}{D-2}(1-\gamma) \Big\} \, .
\eea
Then the field equations (\ref{rfe}) expanded around the background AdS metric convert to $\bar{V}_a = 0$
and $\bar{S} = 0$, respectively. After the expansion of the derivative operator, these two expressions are,
however, still quite cumbersome. In order to proceed with some further simplifications, a simpler and more
efficient approach that we now adopt is to consider the contraction of the field equations with respect to
the vector $k^a$. Taking $\bar{V}_ak^a$ and $\bar{S}$, we find that they can be written as
\bea
\bar{V}_ak^a &=& \gamma S_1 +\frac{\gamma}{D-2}S_2 +(1-\gamma)\bar{\nabla}_a(\bar{D}k^a) \nn \\
&& +2k_a\bar{\nabla}_b\big[(1-\gamma)^2k^{[a}\bar{D}k^{b]}\big] \, , \label{CV} \\
(D-2)\bar{S} &=& e^{\Phi}\big[(1-\gamma)S_1 +(D-2)S_2 \nn \\
&& -2(D-2)(1-\gamma)\bar{\nabla}^a\Phi(\bar{D}k_a) \nn \\
&& +(\bar{D}k)^2\gamma^{-1}(1-\gamma)^3\big] \, , \label{CS}
\eea
in terms of two simple notations
\bea
S_1 &\equiv& 2\big[(\bar{\nabla}^ak^b)\bar{\nabla}_{[b}k_{a]} +g^2(D-3)\big](1-\gamma^{-1}) \nn \\
&& +(D-2)\gamma^{-1}\bar{\nabla}^b[\gamma k_b(\bar{D}\Phi)] \, , \\
S_2 &\equiv& \bar{\square}\Phi -(D-2)(\bar{\nabla}\Phi)^2 \nn \\
&& -2g^2\frac{D-3}{D-2}(1-\gamma^{-1}) \, .
\eea

Considering now the geodesic condition $\bar{D}k^a = 0$ and after substituting it into $\bar{V}_ak^a$ and
$\bar{S}$, one can observe that they are, in fact, only the combinations of two scalars $S_1$ and $S_2$. This
means that the sufficient conditions for the vanishing of the contracted equations $\bar{V}_ak^a$ and the
scalar field equation $\bar{S}$ are that the expressions $S_1$ and $S_2$ must vanish simultaneously as well.
One cannot extract any new useful information other than two simple scalar equations $S_1 = 0$ and $S_2 = 0$.
In particular, note that the condition $S_2 = 0$ is independent of the vector $k^a$, but depends only on the
dilaton scalar $\Phi$.

Having extracted two simple conditions $S_1 = 0$ and $S_2 = 0$ from the contracted field equations (\ref{CV})
and the scalar field equation (\ref{CS}) around the background metric, it is necessary to see whether these
two conditions are also sufficient for the field equation $V_a = 0$ with respect to the background metric.
To this end, we now rewrite the vector expression $\bar{V}_a$ in terms of $S_1$ and $S_2$ as
\be
\bar{V}_a \equiv V_a -\frac{S_2}{D-2}\gamma k_a -2\bar{\nabla}^b\big[(1-\gamma)^2k_{[b}\bar{D}k_{a]}\big]
\ee
where
\bea\begin{aligned}
V_a &\equiv \bar{\nabla}^b\big\{ 2(\gamma-1)\bar{\nabla}_{[b}k_{a]} +(D-2)\gamma k_b\bar{\nabla}_a \Phi\big\} \\
 & -(D-2)\gamma (\bar{\nabla}^b\Phi)\bar{\nabla}_bk_a +2(D-3)(1-\gamma)g^2k_a \, .
\end{aligned}\eea
From the gauge field equation $\bar{V}_a = 0$, one can get a new equation $V_a = 0$. This condition, together
with the geodesic equation $\bar{D}k^a = 0$ and $S_2 = 0$, is sufficient to ensure that the field equation
$\bar{V}_a$ vanishes. It is also worth noting that $S_1$ is simply related to the vector $V_a$ by $S_1 =
\gamma k^aV_a$; thus $V_a = 0$ implies $S_1 = 0$ immediately. This means that $S_1 = 0$ is equivalent to
$\bar{V}_a = 0$.

The remaining step is to consider the Einstein equation (\ref{ree}) and its contraction with the vector $k_a$
once and twice. For this purpose, we first expand it around the background spacetime and convert it to the
form denoted simply as $\bar{E}_{ab} = 0$, where
\bea
\bar{E}_{ab} &\equiv& \frac{1-\gamma^{-1}}{2(D-2)}\big[(D-2)(\gamma +D-3)k_ak_b
 -\bar{g}_{ab}\gamma\big]S_1 \nn \\
&& -(D-2)\gamma \bar{\nabla}_{(a}\Phi\bar{D}k_{b)} +\frac{1}{2(D-2)\gamma^2}\big[(D-2)\gamma S_2 \nn \\
&& +(1-\gamma)^3(\bar{D}k^c)\bar{D}k_c \big]\big[\bar{g}_{ab}\gamma +(\gamma +D-3)k_ak_b\big] \nn \\
&& +(\gamma^{-1}-1)\big[(\gamma -3+D)k_ak_b(\bar{\nabla}^{c}\Phi)\bar{D}k_c \nn \\
&& +(D-2)\gamma(\bar{D}\Phi)\bar{D}(k_ak_b)\big] -T_{ab} \nn \\
&& +(\gamma^{-1}-1)V_{(a}k_{b)} +\frac{(1-\gamma)^2}{2\gamma} \big\{\bar{\nabla}^c[k_c\bar{D}(k_ak_b)] \nn \\
&& +(\gamma-2)(\bar{D}k_a)\bar{D}k_b -(\bar{D}k_d)\bar{\nabla}^d(k_ak_b)\big\} \, ,
\eea
has been recast into its contraction with $k_a$ once and twice, while the non-contractible symmetric part is
\bea
T_{ab} &\equiv& 2(1-\gamma)\big\{g^2(\bar{g}_{ab} +k_ak_b)
 -\bar{\nabla}_{[c}k_{(a]}\bar{\nabla}^{c}k_{b)}\big\} \nn \\
&& +\bar{\nabla}^c\big[(1-\gamma)k_c\bar{\nabla}_{(a}k_{b)}\big] \, .
\eea

As is shown in the above, given the geodesic conditions $\bar{D}k_{a} = 0$, we have obtained three simpler
equations: $V_a = 0$, $S_1 = 0$, and $S_2 = 0$. Using these conditions, a sufficient condition for $\bar{E}_{ab}
= 0$ is that the tensor $T_{ab}$ should vanish as well. Therefore, all the expanded field equations obtained
by the background metric expansion method, $\bar{V}_a = 0$, $\bar{S} = 0$, and $\bar{E}_{ab} = 0$, will be
satisfied if $\bar{D}k_{a} = 0$, $V_a = 0$, $S_2 = 0$, and $T_{ab} = 0$, which have been explicitly verified
with the KK-AdS black hole solutions \cite{SQWu}.

To summarize, we establish that for the stringy Kerr-Schild metrics with a geodesic timelike vector $k^a$, solving
the field equations (\ref{ree}) and (\ref{rfe}) could be reduced to solving straightforwardly the following three
relative simple equations around the background metric:
\bea
&&\hspace*{-8pt} \bar{\square}\Phi -(D-2)(\bar{\nabla}\Phi)^2 -2g^2\frac{D-3}{D-2}(1-\gamma^{-1}) = 0 \, ,
 \qquad \label{S2} \\
&& \bar{\nabla}^b\big\{ 2(\gamma-1)\bar{\nabla}_{[b}k_{a]} +(D-2)\gamma k_b\bar{\nabla}_a \Phi\big\} \nn \\
&&\qquad -(D-2)\gamma (\bar{\nabla}^b\Phi)\bar{\nabla}_bk_a \label{Vb} \\
&&\qquad +2(D-3)(1-\gamma)g^2k_a = 0 \, , \nn \\
&& 2(1-\gamma)\big\{g^2(\bar{g}_{ab} +k_ak_b) -\bar{\nabla}_{[c}k_{(a]}\bar{\nabla}^{c}k_{b)}\big\} \nn \\
&&\quad +\bar{\nabla}^c\big[(1-\gamma)k_c\bar{\nabla}_{(a}k_{b)}\big] = 0 \, . \label{Tab}
\eea
Thus, the sufficient conditions for the sKS ansatz are the geodesic condition (\ref{geoc}) on $k^a$, Eq.
(\ref{S2}) on the dilaton scalar $\Phi$, and the conditions (\ref{Vb}) and (\ref{Tab}) on $k^a$ and $\Phi$.
In particular, the condition (\ref{S2}) depends only on the properties of the dilaton scalar $\Phi$ and
this set of conditions is also satisfied spontaneously in the uncharged case ($\Phi = 0$). In a word, after
assuming that the timelike vector $k^a$ is geodesic, we find that all the field equations with respect to
the background metric then can be reduced to the three conditions $S_2 = 0$, $V_a = 0$, and $T_{ab} = 0$ given
above. Nevertheless, an open question is to see whether these wonderful results derived by the background
metric expansion method for the sKS ansatz would find some as-yet-unknown exact solutions.

\section{Applications: new KK-AdS solutions
with spherical and planar topology}

Inspired by the observation that Eqs. (\ref{S2}), (\ref{Vb}), and (\ref{Tab}) around the background metric
$\bar{g}_{ab}$ can be seen as the counterparts of the field equations (\ref{rfe}), then one wonders naturally
about whether there exists a new vector field that may be different from the known one in the given black
hole solutions but still satisfies all the field equations. As a test of our results derived above, in the
following we shall use the pure AdS background metrics with spherical and planar topology as two concrete
examples to derive new exact solutions.

To present explicitly the general KK-AdS solutions in the below, we shall adopt conventions as those in
\cite{GLPP04,SQWu}. The dimension of spacetime is denoted as $D = 2N +1 +\epsilon \geq 4$, with $N = [(D-1)/2]$
being the number of rotation parameters $a_i$ and $2\epsilon = 1 +(-1)^D$. Let $\Phi_i$ be the $N$ azimuthal
angles in the $N$ orthogonal spatial 2-planes, each with period $2\pi$. The remaining spatial dimensions are
parameterized 
by a radial coordinate $r$ and by $N +\epsilon = n = [D/2]$ ``direction cosines'' $\mu_i$ obey
the constraint $\sum_{i=1}^{N +\epsilon} \mu_i^2 = 1$, where $0\leq \mu_i \leq 1$ for $1\leq i\leq N$, and
$-1\leq \mu_{N+1} \leq 1$, $a_{N+1} = 0$ for even $D$. Moreover, shorthand notations $c = \cosh\delta$ and
$s = \sinh\delta$ are used.

Now we would like to find where there exists a new vector field $K$ tangent to an affinely parameterized
timelike geodesic congruence of the AdS background metric, which is assumed to have the general form
\be
K = K_t (\tilde{\mu}_i)\, dt +K_r(r,\tilde{\mu}_i)\, dr
 +\sum_{i=1}^{N}K_{\phi_i}(\tilde{\mu}_i)\, d\phi_i \, , \label{newK}
\ee
where $K_t$, $K_r$, and $K_{\phi_i}$ are some unknown functions to be specified, and the notation
$\tilde{\mu}_i \equiv \mu_1, \mu_2, ... , \mu_i (i = 1, ... , N+\epsilon)$ has been used.

\subsection{The spherical
AdS background metric}

Consider first the case of a spherical AdS background metric. Supposing that the vector $K$ satisfies the
equations (\ref{Vb}) and (\ref{Tab}) with respect to the pure AdS background metric given in \cite{GLPP04,SQWu}
\bea
d\bar{s}^2 &=& -\big(1+g^2r^2\big)W\, dt^2 +F\, dr^2 \nn \\
 && +\sum_{i=1}^{N +\epsilon}\frac{r^2+a_i^2}{\chi_i}\, d\mu_i^2
 +\sum_{i=1}^N\frac{r^2+a_i^2}{\chi_i}\mu_i^2\, d\phi_i^2 \nn \\
&& -\frac{g^2}{\big(1+g^2r^2\big)W}\bigg( \sum_{i=1}^{N +\epsilon}
 \frac{r^2+a_i^2}{\chi_i} \mu_i\, d\mu_i \bigg)^2 \, , \qquad
\eea
where the scalar functions $W$ and $F$ are
\be
W = \sum_{i=1}^{N +\epsilon}\frac{\mu_i^2}{\chi_i} \, , \quad
F = \frac{r^2}{1+g^2r^2}\sum_{i=1}^{N +\epsilon}\frac{\mu_i^2}{r^2+a_i^2} \, ,
\ee
then the functions $K_t$, $K_r$, and $K_{\Phi_i}$ can be easily calculated. For a comparison with that presented
in \cite{SQWu}, we get the following new vector,
\bea
K &=& \sqrt{c^2C_1 +g^2(\epsilon-1)C_2}W\, dt +\sqrt{f(r)}F\, dr \nn \\
&& -\sum_{i=1}^{N}\frac{\sqrt{a^2_i\Xi_i +(\epsilon-1)C_2}}{\chi_i}\mu^2_i\, d\phi_i
\eea
where $f(r) = c^2C_1 -s^2(1+g^2r^2) -(\epsilon-1)C_2/r^2$, $\Xi_i = c^2C_1 -s^2\chi_i$, and $\chi_i = 1 -g^2a_i^2$,
while $C_1$ and $C_2$ are two arbitrary constants. If $C_1 = 1$ and $C_2 = 0$, then the solution reduces to that
given in \cite{SQWu}. It should be pointed out that we have directly and explicitly checked that the above vector
$K$, together with the full metric and the gauge potential 1-form
\be\begin{split}
ds^2 &= H^{1/(D-2)}\Big(d\bar{s}^2 +\frac{2m}{UH}K^2\Big) \, , \\
A &= \frac{2ms}{UH}K \, , \quad \Phi=\frac{-1}{D-2}\ln(H) \, ,
\end{split}
\ee
obeys the field equations derived from the Lagrangian (\ref{sKSl}) of the Einstein-Maxwell-dilaton theory. In the
above, the scalar functions ($U, H$) are defined to be
\be
U = r^{\epsilon}\sum_{i=1}^{N +\epsilon}\frac{\mu_i^2}{r^2+a_i^2}\prod_{j=1}^N\big(r^2+a_j^2\big) \, , \quad
H = 1 +\frac{2ms^2}{U} \, .
\ee

\subsection{The planar
AdS background metric}

As an input, it has been assumed that the background metric $\bar{g}_{ab}$ is the pure AdS metric; we now
take the planar AdS metric \cite{Awad} as the background spacetime for further verification. Similarly, one
can solve the timelike and geodesic vector $\tilde{K}$ assumed in (\ref{newK}) with respect to the planar
AdS background metric satisfying Eqs. (\ref{Vb}) and (\ref{Tab}). Through a series of tedious calculations,
we obtain the new planar AdS solutions as follows:
\be
\begin{split}
ds^2 =& H^{1/(D-2)}\Big(-g^2r^2\, dt^2 +\frac{dr^2}{g^2r^2} +r^2d\Sigma_k^2 \\
 &+\frac{2ms^2}{r^{D-3}H}\tilde{K}^2 \Big) \\
A =& \frac{2ms^2}{r^{D-3} H}\tilde{K} \, , \qquad \Phi = \frac{-1}{D-2}\ln(H) \, ,
\end{split} \label{PNfullg}
\ee
where $d\Sigma_k^2$ denotes the flat $k = (D -2)$-space unit metric for zero curvature and the timelike
1-form $\tilde{K}$ is given by
\be
\tilde{K} = C_0\, dt +\sum_{i=1}^{N}C_i\, d\phi_i +\sqrt{h(r)}g^{-2}r^{-2}\, dr \, ,
\ee
in which $C_0$ and $C_i$ are some arbitrary $N+1$ constants and the functions ($H, h(r)$) are defined to be
\be\begin{split}
H = 1 +2ms^2r^{3-D} \, , \quad
h(r) = C_0^2 -g^2\Big(r^2 +\sum_{i=1}^{N}C_i^2\Big) \, .
\end{split}\ee

To see whether the solution (\ref{PNfullg}) describes a regular black hole, one can perform the following
coordinate transformations:
\be\begin{split}
dt &\to dt +\frac{2ms^2C_0}{g^2r^{D-1}\Delta}\sqrt{h(r)}\, dr \, , \\
d\phi_i &\to d\phi_i -\frac{2ms^2C_i}{r^{D-1}\Delta}\sqrt{h(r)}\, dr \, .
\end{split}\ee
Then the metric and the gauge potential become
\be\begin{split}
ds^2 =& H^{1/(D-2)}\Big(-g^2r^2\, dt^2 +\frac{dr^2}{\Delta} +r^2d\Sigma_k^2 \\
 & +\frac{2ms^2}{r^{D-3}H}K^2 \Big) \, , \\
A =& \frac{2ms^2}{r^{D-3}H}K \, , \qquad K = C_0\, dt +\sum_{i=1}^{N}C_i\, d\phi_i
\end{split}\ee
where
\be
\Delta = g^2r^2 -2ms^2r^{3-D}h(r) \, .
\ee
The horizon is determined by $\Delta = 0$ and is endowed with a planar topology.

\section{Conclusions}

In this paper, we have studied a new metric ansatz dubbed as the stringy Kerr-Schild ansatz since it can
been seen as the most meaningful generalization of the Kerr-Schild form for the (un)gauged supergravity
theory, in which the general black hole solutions in all dimensions share a common and universal metric
structure.

Initially, we have adopted the traditional method of perturbation expansion for this new ansatz, because
it had already been successfully applied in the extended KS ansatz \cite{xKS}. We attempted to explore
some properties of the sKS form in the usual way. But unfortunately, as a consequence, the introduction
of the perturbation factor appears clearly to be in contradiction with the basic assumption of the sKS
ansatz. Therefore, the traditional perturbation expansion analysis of the full metric tensor provides no
help to reach our aim.

As such, we have proposed a new method of background metric expansion to extract simple information by
expanding the field equations of the full spacetimes around the background metric. In Sec. III, we
obtained the geodesic condition (\ref{geoc}) from the Lagrangian of the Einstein-Maxwell-dilaton system
with respect to the background metric and the counterparts (\ref{S2}), (\ref{Vb}), and (\ref{Tab}) of all
the field equations around the background metric. The condition (\ref{S2}) depends only on the properties
of the dilaton scalar $\Phi$, and what is more, the set of the sufficient conditions (\ref{S2}), (\ref{Vb}),
and (\ref{Tab}) is satisfied spontaneously in the uncharged case; thus our method coincides with the usual
method of perturbation expansion for the Kerr-Schild form. As anticipated, the overall calculations can be
substantially simplified in our method; meanwhile, the results of our analysis could be helpful in
obtaining new exact solutions of the sKS form. As two examples of applications of our method, we first
rederived the rotating single-charged KK-AdS solutions by further introducing one or two arbitrary constants
(corresponding to even and odd dimensions, respectively). Moreover, for further verification, we obtained
new solutions by using the planar anti-de Sitter metric as the background one in sKS ansatz.

It is worthwhile to investigate whether the analysis made in this paper can be generalized to multiple-charged
black hole solutions \cite{Wu2c,Wu5d3c} in supergravity theories, since the general nonextremal rotating
charged AdS black hole solutions with two independent charge parameters still remain elusive in $D = 6, 7$
gauged supergravities.

\begin{acknowledgments}
This work is supported by the NSFC under Grant No. 10975058 and No. 11275157.
H. Wang is grateful to Prof. H. L\"{u} for helpful discussions and comments.
\end{acknowledgments}

\medskip
\appendix
\section{Double Kerr-Schild form for $D=4$ Kerr-NUT-AdS solution}

The four-dimensional Kerr-NUT-AdS metric admits a double Kerr-Schild representation as follows:
\begin{widetext}
\bea
ds^2 &=& -\frac{(1+g^2r^2)(1-g^2y^2)}{\chi}\, dt^2 +\frac{(r^2+y^2)\, dr^2}{(r^2+a^2)(1+g^2r^2)}
 +\frac{(r^2+y^2)\, dy^2}{(a^2-y^2)(1-g^2y^2)} +\frac{(r^2+a^2)(a^2-y^2)}{a^2\chi}\, d\phi^2 \nn \\
&& +\frac{2mr}{r^2+y^2}K^2 +\frac{2ny}{r^2+y^2}N^2 \, , \nn
\eea
where two null 1-forms are
\be
K = \frac{1-g^2y^2}{\chi}\, dt -\frac{a^2-y^2}{a\chi}\, d\phi
 -\frac{(r^2+y^2)\, dr}{(r^2+a^2)(1+g^2r^2)} \, , \qquad
N = \frac{1+g^2r^2}{\chi}\, dt -\frac{r^2+a^2}{a\chi}\, d\phi
 -\frac{i(r^2+y^2)\, dy}{(a^2-y^2)(1-g^2y^2)} \, . \nn
\ee
\end{widetext}
The vectors $K_{\mu}$ and $N_{\mu}$ are two \emph{linearly independent}, mutually orthogonal, affinely-parameterized
null geodesic congruences, so they need not be proportional to each other. Note that $N_{\mu}$ is a complex vector
rather than a real vector.

The most general Pleba\'nski-Demia\'nski type-D solution \cite{Plebanski2} with an extra acceleration parameter
can be put into a similar form. Higher-dimensional generalizations with just one rotation parameter were presented
in Ref. \cite{doubleKS}, while the multi-Kerr-Schild form \cite{mulKS} has been studied in detail for the most
general AdS solution with NUT charges.

\end{document}